# Second-Order Synaptic Memory Using Inherent Plasticity of Moiré Superlattices

*Tanweer Ahmed,[1]* Kenji Watanabe,[2] Takashi Taniguchi,[3] Fèlix Casanova,[1,4] and Luis E. Hueso [1,4]**

[1]*CIC nanoGUNE, BRTA, 20018 Donostia-San Sebastian, Basque Country, Spain.*
[2]*Research Center for Electronic and Optical Materials, National Institute for Materials Science, 1-1 Namiki, Tsukuba 305-0044, Japan.*
[3]*Research Center for Materials Nanoarchitectonics, National Institute for Materials Science, 1-1 Namiki, Tsukuba 305-0044, Japan.*
[4]*IKERBASQUE, Basque Foundation for Science, 48009 Bilbao, Basque Country, Spain.*

Keywords: Twisted double bilayer graphene, second-order nonlinear electrical response, extrinsic mechanisms, synaptic memory, moiré superlattices.

E-mail: t.ahmed@nanogune.eu, l.hueso@nanogune.eu

Achieving synaptic functionality electronically in a single-element quantum material is a fundamental challenge, as conventional methods rely on the introduction of extrinsic charge-traps or polar components. Here, we demonstrate that twisted double bilayer graphene (tDBLG) moiré superlattices—composed purely of carbon—exhibit electronic hysteresis and plasticity in presence of twist-angle disorder. Inversion symmetry breaking at the moiré length scales also gives rise to second-order nonlinear electrical response via disorder-mediated extrinsic mechanisms. Such second-order nonlinearity is highly tunable in both sign and magnitude by varying carrier concentration and vertical displacement field. We harness the coexistence of electronic plasticity and second-order nonlinearity to realize a second-order synaptic memory device. Our findings establish strained moiré carbon systems as a powerful new platform for energy-efficient neuromorphic computing, demonstrating that complex electronic functionality can emerge purely from symmetry-breaking physics in a single-element material.

## 1. Introduction

Disorder, an inherent characteristic of materials across all scales, plays a critical role in shaping their electronic properties [1]. The ability to control and exploit disorder has been pivotal to ongoing technological advancements, including the silicon electronics revolution [2,3] and quantum metrology [4,5]. In two-dimensional (2D) van der Waals (vdW) crystals, disorders manifest ubiquitously, stemming from both defects in synthetic bulk crystals as well as impurities introduced during fabrication [6]. In the past few years, moiré superlattices of vdW crystals have emerged as versatile platforms for exploring exotic quantum states, such as correlated superconductivity [7,8], magnetic [9] and ferroelectric orders [10–13], and quantum geometric effects [14,15]. These systems exhibit a distinct form of disorder: interfacial strain fields [16–18] locally modify the twist-angle [15,19–24], leading to inhomogeneous variations of energy and length scales of moiré patterns [25], also known as twist-angle anomaly. Additionally, strain and strain gradients [26] give rise to piezoelectricity [27–30] and flexoelectricity [29–31], transforming this disorder into a source of emergent functionalities [21]. The challenge now lies in leveraging these disorder-engineered moiré systems to develop practical devices by mastering the interplay between imperfections and quantum phenomena for next-generation technologies.

The recent progress in second-order nonlinear electrical responses (NLER) in materials lacking inversion symmetry [14,32–40] has unlocked transformative opportunities [37,41–43]. NLER arise from a combination of disorder-mediated extrinsic effects [33] and intrinsic mechanisms rooted in quantum geometry [34,38]. Unlike conventional nonlinear devices such as diodes, which rely on junctions or interfaces, NLER originate directly from the bulk properties of materials, offering a new paradigm for memory and energy harvesting technologies [37,41,42,44,45]. Moiré superlattices, with their remarkable tunability and pronounced nonlinear responses [14,35,36,46], are at the forefront of this field. Here, flat bands and Lifshitz transitions can be precisely manipulated by controlling the interlayer twist-angle [25,47,48]. NLER are often pronounced near these critical points of the band structure [14,32]. Graphene-based moiré superlattices are particularly notable for their broadband nonlinear response [37,49,50]. Additionally, strain and disorder interact with electronic states via flexoelectric mechanisms, introducing electric field-dependent hysteresis [15,31]. This coupling provides a unique opportunity to integrate electronic hysteresis with band-structure-dependent nonlinear responses, paving the way for efficient second-order memory devices. However, there is a still gap in understanding the impact of strain-induced hysteresis in graphene-based moiré systems for second-order memory applications, addressing which is essential for developing a new breed of synaptic devices based on single elemental materials.

Here we perform first and second-order transport measurements in low-angle ( $\theta$ ) twisted moiré superlattices of double bilayer graphene (tDBLG) [14,15,35,36], hosting the interfacial strain-induced twist-angle anomaly [21]. Our strained tDBLG devices display robust electronic plasticity, characterized by a hysteretic evolution of the electronic band structure under a vertical displacement field, accompanied by a vertical polarization of approximately $0.13\,\mu$C cm$^{-2}$. Additionally, we observe a second-order NLER from our samples, due to inversion symmetry breaking at the moiré length scales. Both the sign and the magnitude of the NLER are highly tunable, particularly close to the integer fillings of the superlattice bands. We combine the tunable electronic plasticity and the NLER to demonstrate a second-order synaptic memory device important for low-power-consuming neuromorphic computation. Our work demonstrates that synaptic functionality can be achieved in strained graphene-based moiré superlattices, without needing polar components [12,13] or charge trapping layers [51,52].

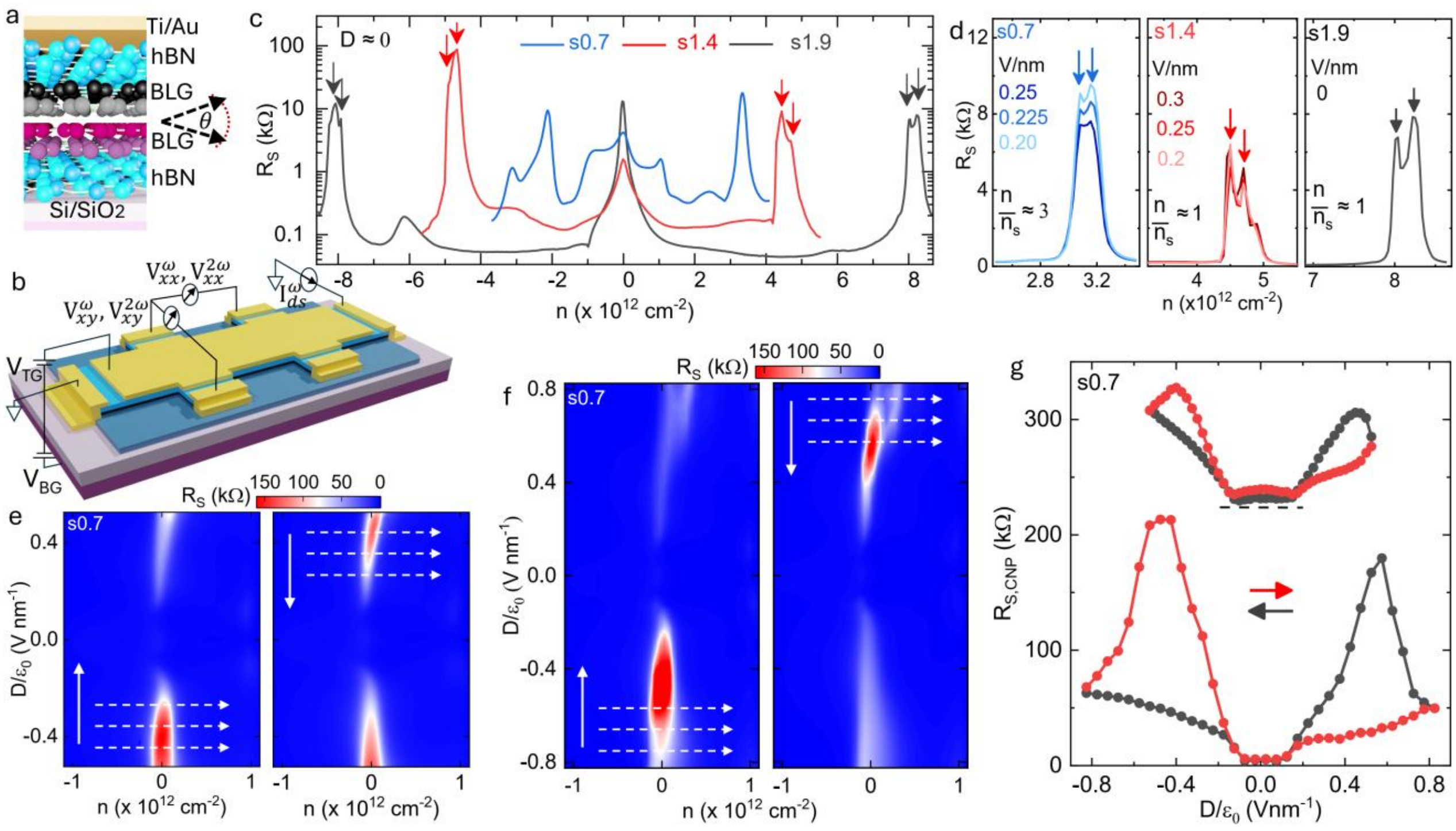

Figure 1: ***D* dependent hysteresis in strained tDBLG superlattice: a,** the cross-section of the stacking sequence of tDBLG devices is schematically shown, the twist angle ($\theta$) is indicated. **b,** the measurement geometry is schematically illustrated. Longitudinal ($V_{xx}$) and transverse ($V_{xy}$) voltages in both the first ($\omega$) and second ($2\omega$) harmonics are measured simultaneously under applied current bias ($I^{\omega}$). **c,** sheet resistance ($R_S$) *vs.* carrier density ($n$) characteristics at vertical fixed displacement field ($D$) $\approx 0$ from three tDBLG samples are presented. $R_s$ peak at integer superlattice filling ($n/n_s$) appears to be a combination of at least two $R_S$ maxima (labeled with arrows). This indicates twist-angle anomaly, arising from interfacial strain. **d,** $R_s\ vs.\ n$ data from three different samples, zooming in close to integer $n/n_S$ for different values of $D/\varepsilon_0$. **e** and **f,** $n\ vs.\ D$ phase spaces of $R_S$ close to CNP are shown for two $D$ sweep ranges from sample s0.7. $n$ is swept (horizontal arrows) at fixed value of $D$. $D$ is gradually changed (vertical arrows) after each $n$ sweep. Phase spaces collected with ascending and descending $D$ sweep directions confirm presence of $D$-dependent hysteresis of $R_S$. **g,** $R_S$ at CNP ($R_{S,CNP}$) vs. $D$ hysteresis from sample s0.7 is shown for two different sweep ranges (vertically shifted, horizontal dashed line is the baseline). $D$ sweep directions are indicated (red and black arrows).

## 2. Results and discussion

The dual-gated hexagonal boron nitride (hBN)-encapsulated AB-AB stacked low-angle ($\theta$) twisted double bilayer graphene (tDBLG) field effect transistors (FETs) were fabricated using usual dry transfer [53,54] and tear-and-stack method [7]. We studied six samples, namely s0.7, s1.3, s1.4, s1.4(2), s1.5, and s1.9 hosting low $\theta \approx 0.7°, 1.3°, 1.4°$, $1.4°$, $1.5°$, and $1.9°$, respectively [24,47,55–57]. For comparison, we also characterized a control sample (s10) hosting a larger $\theta \approx 10°$. The details of the fabrication and samples are provided in the supplementary information section 1. A schematic of the cross-section of the sample is shown in Figure 1a. The dual gated tDBLG FET with the measurement circuit configuration is schematically presented in Figure 1b. Here, we independently control the carrier density ($n$) and vertical displacement field ($D$), by simultaneously sweeping both top ($V_{tg}$) and bottom gate ($V_{bg}$) voltages (See methods). We simultaneously measure voltage drops in both first ($\omega$) and second ($2\omega$) harmonics in longitudinal ($V_{xx}^{\omega}, V_{xx}^{2\omega}$) and transverse ($V_{xy}^{\omega}, V_{xy}^{2\omega}$) direction in presence of applied current bias ($I_{ds}^{\omega}$) of $\omega$ frequency using synchronized lock-in amplifiers. The sheet resistance ($R_S$) vs. $n$ characteristics at $D \approx 0$ from samples s0.7, s1.4, and s1.9 are presented in Figure 1c. We observe strong $R_S$ maxima corresponding to the full filling ($n = n_s$) of the first superlattice band at $n \approx \pm 4.5 \times 10^{12}$ and $n \approx \pm 9.5 \times 10^{12}$ cm$^{-2}$, for samples s1.4 and s1.9, respectively. For sample s0.7, we observe $R_S$ maxima at integer multiples of $n \approx \pm 1.1 \times 10^{12}$ cm$^{-2}$. We calculate $\theta$ using the relation $n_s = 4/A$, [7] where $A \approx \sqrt{3}a^2/(2\theta^2)$ is the area of the moiré unit cell and $a$ is the lattice parameter of graphene. Interestingly, the $R_S$ peaks at the integer $n/n_s$ are observed to be a smeared-out combination of at least two distinct $R_S$ maxima, marked with downward arrows. Such splitting of the $R_S$ maxima is shown in the three panels of Figure 1d, focusing close to integer $n/n_s$. See SI section 3 for the data from sample s1.3. Our samples are clean enough to demonstrate fully (both spin and valley) symmetry-broken Landau levels in [55]. In the Landau fan data, we also observe two parallel-dispersing Landau fans originating from doubly-split $R_S$ maxima (see SI section 4) at $n/n_s = 1$. Furthermore, the splitting of $R_S$ maxima is observed in all values of $D$, and the doubly-split $R_S$ maxima disperse in parallel in the $D$ vs. $n$ phase space (see SI section 5). These observations confirm the presence of two distinct values of $n_s$ resulting from two different moiré lengths, appearing from twist-angle anomaly [24,15,21], a manifestation of interfacial strain [16–18]. This form of disorder is commonly observed and difficult to avoid in moiré superlattices. Raman spectroscopic characterization on our samples reveals up to ~ 0.05% strain variations in the graphene channel, which is presented in the SI section 2.

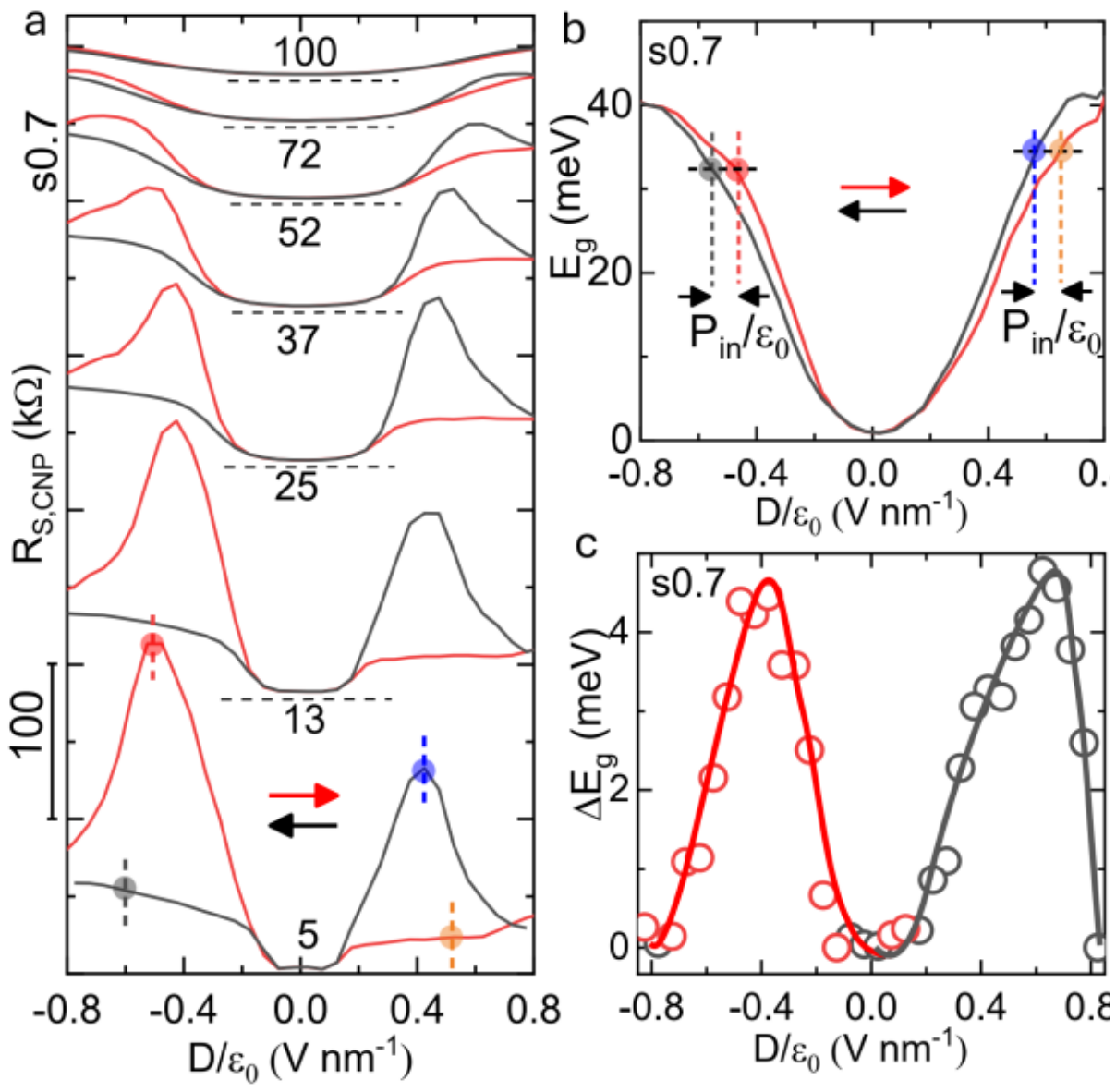

Figure 2: ***D*** **-dependent hysteresis of the band structure: a,** the $R_{S,CNP} - D$ hysteresis loops collected from sample s0.7 at different $T$, which are vertically shifted. The baselines are shown as dashed lines. red (black) traces present the ascending (descending) $D$ sweep directions (indicated with arrows). **b,** estimated band-gap ($E_g$) vs. $D$ data for both sweep direction is presented. The hysteresis in the $E_g - D/\varepsilon_0$ characteristics indicates the appearance of an internal vertical polarization ($P_{in}$). The red (orange) and black (blue) circles represent the same value of $E_g$, for ascending and descending $D$ sweeps, respectively, However, they exhibit significantly different values of $R_{S,CNP}$. **c,** $\Delta E_g$ between the two states are presented as a function of $D/\varepsilon_0$. The red (black) points indicate higher $E_g$ for the ascending (descending) sweep directions.

After confirming the presence of a twist-angle anomaly, we now focus on the dependence of $R_S$ on $D$ and $n$, close to $n/n_s = 0$ or the charge neutrality point (CNP) from all the samples. Figure 1e demonstrate the $n - D$ phase space of $R_S$ from s0.7 at temperature $T = 2$K. Here, $n$ is the fast axis, and $D/\varepsilon_0$ is the slow axis. $D$ is kept fixed while n is varied in small steps. The $n$ sweeps are marked as horizontal dashed arrows. After every $n$ sweep $D/\varepsilon_0$ is changed by a small step and then $n$ is varied again. The left and right panels show the data for ascending and descending $D$ sweep directions, respectively, which are indicated with vertical arrows. We observe a striking $D$ dependent hysteresis, which becomes pronounced upon increasing the sweep range of $D$ (Figure 1f). Figure 1g demonstrates the hysteretic $R_{S,CNP}$ ($R_S$ at CNP) vs. $D$ data for two different $D$ sweep ranges, which are vertically shifted for clarity. The baseline is indicated as a dashed horizontal line. The $D$ sweep directions are indicated with arrows. Such hysteresis is observed in all samples (See SI section 6 for the data of sample s1.3, s1.4, and s1.9), while the rest of main text focuses on the data from sample s0.7. Apart from the hysteresis, $R_S$ dramatically increases beyond $|D/\varepsilon_0| > 0.15$ V nm$^{-1}$, marking the well-known topological transition in the tDBLG band structure [47,56]. The observed butterfly-like hysteresis has been previously reported in 2D vdW ferroelectrics [58]. However, in AB-AB stacked tDBLG, the appearance of ferroelectricity is unlikely. Here, the moiré lattice is stabilized by creating ABAB, ABBC, ABCA stacking orders [59]. Although the moiré lattice is chiral and inversion symmetry is broken at the moiré length scale, the local stacking orders lack inversion symmetry breaking, forbidding ferroelectricity. Interestingly, we do not observe a strong shift in the CNP for ascending and descending sweep directions. Such hysteresis is extremely stable under multiple $D/\varepsilon_0$ sweep cycles (See SI section 7). Furthermore, the hysteresis can be precisely controlled by adjusting the $D$ sweep range (See SI section 8). Such hysteresis is prominently observed not only at CNP but also higher superlattice fillings ($n/n_s = \pm 1, \pm 2, \pm 3$, etc, See SI section 9).

To understand the origin of such hysteresis, we performed the $n$ vs. $D$ phase space measurements of $R_S$ as a function of the temperature ($T$), close to the CNP. We present the $R_{S,CNP} - D$ hysteresis from sample s0.7 at different values of $T$ in Figure 2a (See SI section 10 for data from s1.9). The sweep directions of $D$ are indicated using the arrows. The data at different $T$ are vertically shifted. The horizontal dashed baselines are indicated. Hysteresis is strongest at the lowest $T$ and gradually reduces with increasing $T$. We determine the band-gap ($E_g$) from the Arrhenius relation of $R_{S,CNP}$ ($2k_B \ln R_{S,CNP} \propto E_g/T$) at fixed values of $D/\varepsilon_0$ [55], here $k_B$ is the Boltzmann constant. See SI section 10 for more details. The $E_g$ vs. $D/\varepsilon_0$ data for both sweep directions are plotted in Figure 2b. The data indicates a hysteretic evolution of the electronic band-structure. Figure 2c presents the absolute value of the difference in the $E_g$ between ascending and descending sweep directions. The red (black) data points indicate that the $E_g$ is higher for ascending (descending) sweep direction. The solid lines are the guide to the eye. At intermediate values of $D$ ($\approx \pm 0.4$Vnm$^{-1}$), the $E_g - D$ characteristics appear to be horizontally shifted by $P_{in}/\varepsilon_0 \sim 0.15$ Vnm$^{-1}$ (Figure 2b). Here $P_{in}$ ($\approx 0.13\ \mu$Ccm$^{-2}$) is the effective vertical polarization which constructively, and destructively, contributes to the applied vertical $D$, during descending and ascending sweep of $D$, respectively, similar to a ferroelectric polarization. However, unlike ferroelectric polarization reported in the moiré superlattice of BLG [10–13], our $P_{in}$ is unsaturated and becomes small at $D \approx 0$, instead of remaining constant. Such striking differences in the observed hysteresis raise the question regarding its origin. This also cannot be explained by the presence of interfacial traps, as confirmed by the absence of such hysteresis in a control BLG sample (see SI section 11).

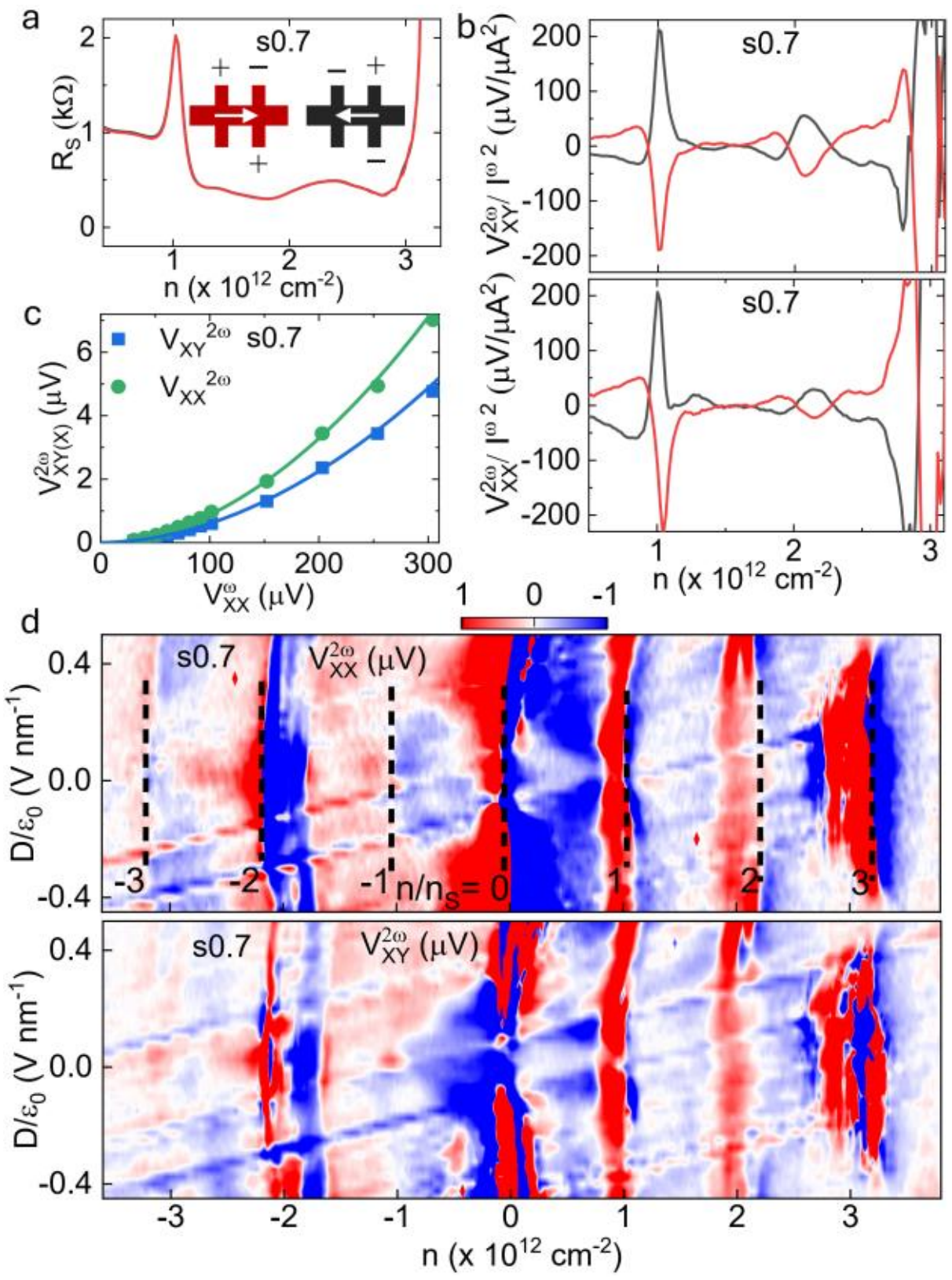

Figure 3**: Second-order nonlinear electrical response from tDBLG: a,** The black and red traces are $R_S - n$ data from sample s0.7, measured in two measurement configurations with opposite polarities of current and voltage probes (black and red schematics in the insets). **b,** The top and bottom panels show the $V_{xy}^{2\omega}/I_{ds}^{\omega\,2}$ and $V_{xx}^{2\omega}/I_{ds}^{\omega\,2}$ vs. $n$ data at $D/\varepsilon_0 = 0$, respectively, collected in two measurement configurations. The data demonstrate a clear sign change, upon changing the configuration. **c,** The $V_{ds}^{\omega}$ dependence of $V_{xx}^{2\omega}$ and $V_{xy}^{2\omega}$ at $D/\varepsilon_0 = 0.125$ V nm$^{-1}$ at $n \approx 0.5\, n_s$. The solid lines present the parabolic fit ($V_{xy(x)}^{2\omega} \propto V_{ds}^{\omega\,2}$). **d,** The top and the bottom panels show $D/\varepsilon_0$ vs. $n$ phase spaces of $V_{xx}^{2\omega}$ and $V_{xy}^{2\omega}$, respectively. The vertical dashed lines indicate the sign change in $V_{xx}^{2\omega}$ close the integer fillings ($n/n_s = 0, \pm1, \pm2, \pm3$, etc.) of superlattice bands. Similar sign changes are also observed in the $V_{xy}^{2\omega}$ data.

Recently, unconventional ferroelectric hysteresis in BLG/hBN and SLG/hBN moiré superlattices has been reported when the crystallographic axes of top and bottom hBN are aligned [10–13], creating inversion asymmetric stacking order. In such cases, a strong change in doping (shift in CNP along $n$ axis) is observed for ascending and descending $D$-sweep directions — a behaviour notably absent in our work (see SI section 12). Here we investigate hysteresis in tDBLG moiré superlattices without intentional alignment of the encapsulating hBN layers. In this context, $D$-dependent hysteresis of $R_S$ can arise through two mechanisms: (i) metastable polarized states induced by a layer- dependent doping profile in tDBLG [15], and (ii) a flexoelectric effect [15,31]. The mechanism (i) can exist at all values of $\theta$ [15]. Remarkably, contrary to this, we did not observe the hysteresis in sample s10, a control tDBLG sample with a much larger $\theta \approx 10°$(see SI section 11). On the other hand, mechanism (ii) relies on the presence of strain and strain gradients, which become increasingly impactful at low $\theta$ [60]. Moiré superlattices host significant intrinsic strain and strain gradients near solitonic domain walls [19,26]. Additionally, uncontrolled interfacial strains lead to a spatially varying twist-angle anomaly, creating an inhomogeneous distribution of these domain walls and strain gradients [19–21,31]. In our samples, vertical displacement field can couple with inhomogeneous strain gradients through the flexoelectric mechanism [29–31]. Furthermore, electromechanical breathing of the strained domain walls are pinned by disorder leading to a hysteretic behaviour [15,61–63]. We observe a dramatic difference in $R_{S,CNP}$ values in the ascending and descending sweep directions of $D$, at the same value of the $E_g$. For example, the data marked black and red (blue and orange) circles the Figure 2b and 2a have the same value of $E_g$. This indicates a spatially non-uniform $P_{in}$ and points towards inhomogeneous flexoelectric coupling as its origin. Although the exact nature of the strain gradients is difficult to determine using mesoscopic transport experiments, we focus on exploiting this robust and controllable hysteresis for multifunctional applications. Simultaneous low temperature transport and scanning probe measurements will be needed to pin-point the fundamental origin of the hysteresis.

After discussing the hysteresis, we now focus on the second-order nonlinear electrical response (NLER) which is inherent to any inversion symmetry broken system [14,15,32–38,41,42]. Inversion symmetry is broken in tDBLG in the moiré length scale, perfect for emergence of NLER and its applications. The second-order current density $j_a^{2\omega} = \sigma_{abc}^{2\omega} E_b^{\omega} E_c^{\omega}$ describes the material's nonlinear response to an applied electric field $E^{\omega}$, where $\sigma_{abc}^{2\omega}$ is a third-rank tensor representing the intrinsic second-order conductivity, a property dictated by the material's symmetry and electronic structure, independent of system dimensions or field strength. $a, b, c$ are indices the corresponding to spatial axes (e.g. $x$, $y$, $z$). Here, $V_{xy(x)}^{2\omega}$ (measured using geometry depicted in Figure 1b) is related to $\sigma_{y(x)xx}^{2\omega}$ via the relation $V_{xx}^{2\omega} = \sigma_{xxx}^{2\omega} \frac{{V_{xx}^{\omega}}^2}{\sigma L}$ ($V_{xy}^{2\omega} = \sigma_{yxx}^{2\omega} \frac{W{V_{xx}^{\omega}}^2}{\sigma L^2}$). $L$ and $W$ are the length and width of the sample. $\sigma$ is the first order conductivity. The red and black traces in Figure 3a, present the $R_S - n$ characteristics from sample s0.7, measured in two measurement configurations with opposite current and voltage probes polarities. The two configurations are indicated as black and red schematics in the insets. The data is taken at $T = 2\text{K}$ using $I_{ds}^{\omega} = 200$ nA, and at $D = 0$. Of course, the red and the black traces collapse on top of one another. The top and the bottom panels of Fig 3b present the $V_{xy}^{2\omega}$ and $V_{xx}^{2\omega}$ vs. $n$ characteristics for both measurement configurations, respectively. The sign of the $V_{xy(x)}^{2\omega}$ changes upon switching the measurement configurations. Moreover, we measure $V_{xy(x)}^{2\omega}$ at different values of $I_{ds}^{\omega}$ ranging from 30 nA to 300 nA. The $V_{xy(x)}^{2\omega}$vs. $V_{xx}^{\omega}$ data at $D = 0$ and $n = 0.5\ n_s$ is shown in Figure 3c. The solid lines present the $V_{xy(x)}^{2\omega} \propto {V_{xx}^{\omega}}^2$ fits. Such parabolic dependence and the sign change in $V_{xy(x)}^{2\omega}$ upon changing the measurement configuration conclusively proves the NLER as the origin of the second-order response. The top and the bottom panels of the Figure 3d present the $D$ vs. $n$ phase spaces of $V_{xx}^{2\omega}$ and $V_{xy}^{2\omega}$, respectively taken with $I_{ds}^{\omega} = 100$ nA (see SI section 5 for $D$ vs. $n$ data of $R_S$). Both of $V_{xx}^{2\omega}$ and $V_{xy}^{2\omega}$ have similar order of magnitude, suggesting disorder mediated extrinsic mechanism as their origin, which is further confirmed by the temperature dependent scaling behaviour [32,33,40] of NLER (see SI section 13 for more details). The sign change of NLER [32] at integer superlattice band fillings are marked with vertical dashed lines. Such NLER, when combined with the tunable hysteresis of the tDBLG superlattices, offers a promising approach for low-power multifunctional memory application, where both the sign and magnitude of the NLER memory levels can be deterministically chosen by fixing $n$. Additionally, NLER does not necessarily require a source-drain power, as an efficient NLER can harvest power from an electromagnetic radiation [37,41,42,49,50]. Moreover, the graphene-based materials are theoretically predicted to demonstrate NLER in a wide range of frequencies ranging from DC to THz [37,49,50].

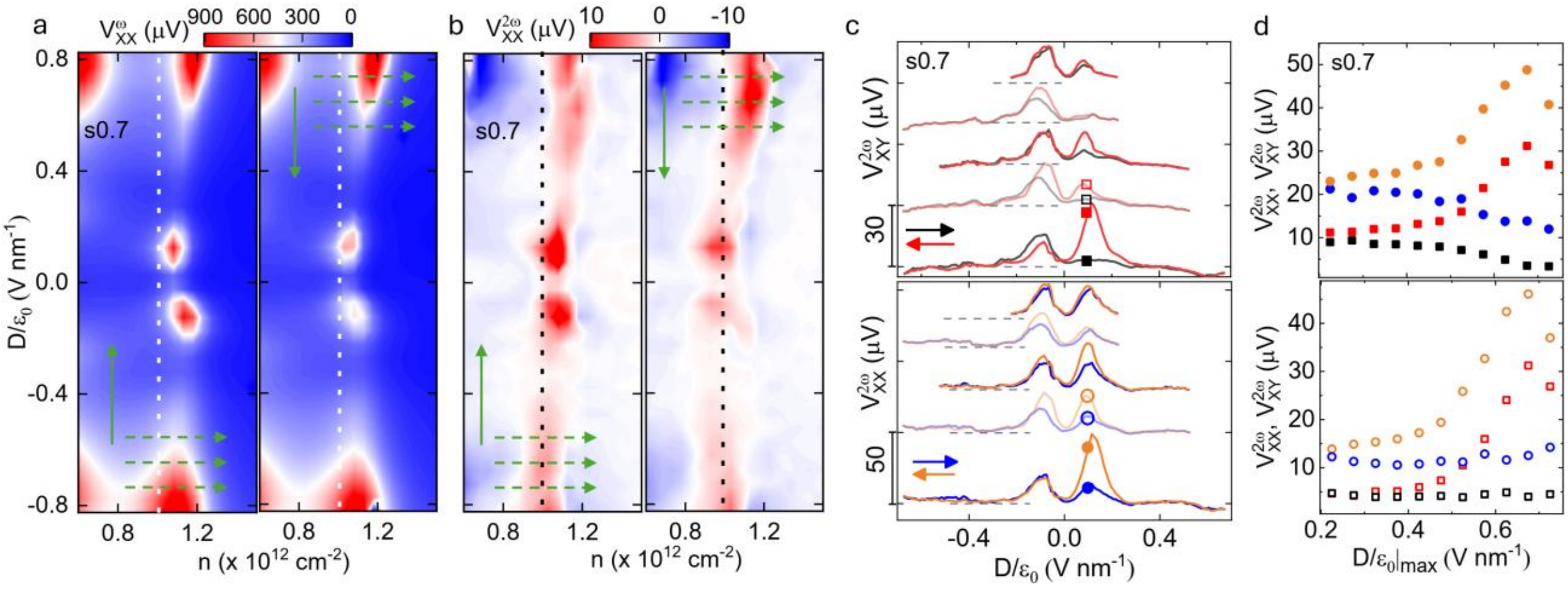

Figure 4: **Second-order plasticity of tDBLG: a** and **b,** $n$ vs. $D/\varepsilon_0$ phase spaces of $V_{xx}^{\omega}$ and $V_{xx}^{2\omega}$, respectively, close to $n/n_s = 1$ is shown from sample s0.7. Sweep directions are marked with green arrows. The synaptic memory action was executed along the vertical dashed line. **c,** The top and the bottom panels present the sweep range dependent hysteresis $V_{xx}^{2\omega}$ and $V_{xy}^{2\omega}$, respectively, for a few different $D$ sweep range. The data of different sweep ranges are vertically shifted. The sweep directions are indicated with arrows. The light and dark traces indicate data for symmetric and asymmetric sweep ranges. The low and the high states (at $D/\varepsilon_0 = 0.1$ Vnm$^{-1}$) are marked. Also see SI section 15 and 17 for more hysteresis data with varying $D$ sweep range in smaller steps, from sample s0.7 and s1.4, respectively. **d,** The values for low (blue and black points) and high (orange and red points) states of $V_{xx}^{2\omega}$ (circles) and $V_{xy}^{2\omega}$ (square) are presented as a function of $D/\varepsilon_0|_{max}$, *i.e.*, for different sweep ranges. The top and the bottom panels presents the data for symmetric and asymmetric sweep ranges.

After discussing the tunable hysteresis and NLER, we combine the two to demonstrate a multifunctional memory with synaptic functionality [51]. We first discuss the $D$ dependent hysteresis at $n \approx 10^{12}$cm$^{-2}$ (from sample s0.7). Figure 4a presents the $n - D$ phase spaces of the $V_{xx}^{\omega}$ measured at $I_{ds}^{\omega} = 100$ nA, close to $n/n_s = 1$. See SI section 5 for full phase space. Faster $n$ sweeps and slow $D$ sweeps are indicated with dashed horizontal, and solid vertical arrows, respectively. The left and right panels show the data for ascending and descending $D$ sweep directions. The simultaneously acquired data of $V_{xx}^{2\omega}$ is presented in the Figure 4b. See SI section 14 for the simultaneously acquired data of $V_{xy}^{2\omega}$. The synaptic device operation was performed along the vertical dashed line. The $D$ dependent hysteresis in $V_{xx}^{\omega}$ and $V_{xx(y)}^{2\omega}$, at $n/n_s \approx 1$, is assisted by the change in the magnitude of the signal ($V_{xx}^{\omega}$and $V_{xx(y)}^{2\omega}$) and also by the small shift in CNP along the $n$ axis (see SI section 12). The sweep range-dependent hysteresis of $V_{xy}^{2\omega}$ and $V_{xx}^{2\omega}$ (using $I_{ds}^{\omega} = 200$ nA) are presented in the top and the bottom panels of Figure 4c, respectively. The dark (light) colours in Figure 4c show the hysteresis using symmetric (asymmetric) ranges of $D$, only for a few sweep ranges, namely − 0.675 to + 0.675

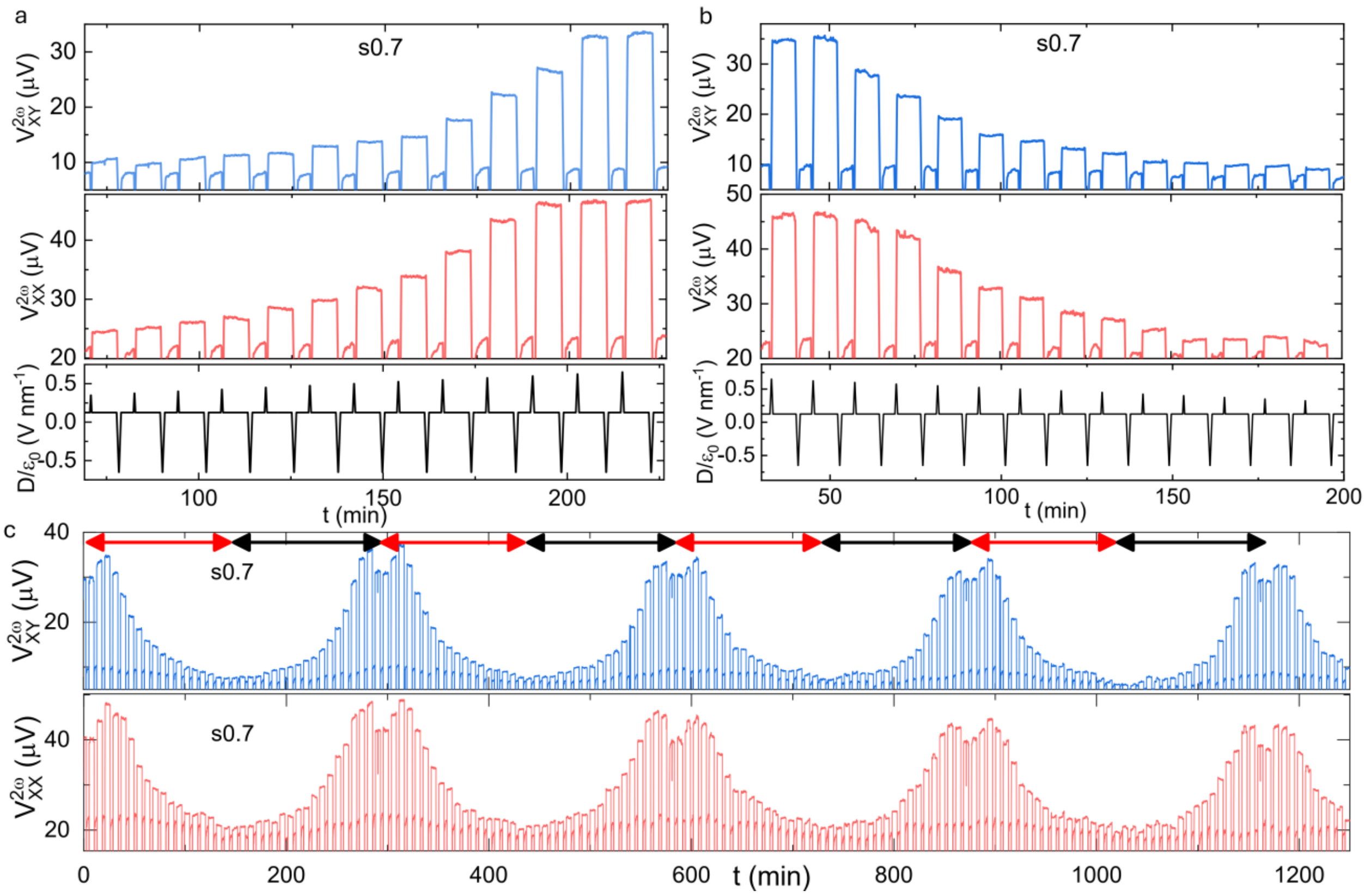

Figure 5: **Second-order synaptic memory operation: a** and **b,** The depression and potentiation cycles of the second order synaptic response is presented, respectively. The timeseries of the sequence of negative (constant) and positive (of varying magnitude) triangular pulsed pulses are shown in the bottom panels. The top and the middle panels present the synaptic response of $V_{xy}^{2\omega}$ and $V_{xx}^{2\omega}$, respectively. **c,** the long-term synaptic plasticity of the second-order response ($V_{xx}^{2\omega}$ and $V_{xy}^{2\omega}$) over several potentiation (red arrows) and depression cycles are presented (black arrows). Also see SI section 17 for synaptic operation in sample s1.4.

Vnm$^{-1}$, – 0.525 to + 0.525 Vnm$^{-1}$ (– 0.675 to + 0.525 Vnm$^{-1}$) and – 0.225 to + 0.225 Vnm$^{-1}$ (– 0.675 to + 0.225 Vnm$^{-1}$). Data for individual sweep ranges have been vertically shifted. The baselines are indicated with horizontal dashed lines for clarity. See SI section 15 for data on $V_{xx}^{\omega}$, $V_{xx}^{2\omega}$ and $V_{xy}^{2\omega}$ at various intermediate sweep ranges. The high and low states of $V_{xy}^{2\omega}(V_{xx}^{2\omega})$, at $D/\varepsilon_0 =$ 0.1 Vnm$^{-1}$, are marked with red (orange) and black (blue) markers, respectively. Interestingly, the magnitude of the high state gradually reduces when the sweep range is reduced. The top panel of Figure 4d presents the values of the $V_{xx}^{2\omega}$ (circular points) and $V_{xy}^{2\omega}$ (square points) at $D/\varepsilon_0 = 0.1$V nm$^{-1}$ at the high (orange and red points) and low states, (blue and black data points), for different symmetric sweep range as a function of $D/\varepsilon_0|_{max}$. Here, the $D/\varepsilon_0$ was swept back and forth between $D/\varepsilon_0|_{max}$ and $D/\varepsilon_0|_{min}$ $(= -D/\varepsilon_0|_{max})$. The bottom panel of the Figure 4d presents the values of the $V_{xx}^{2\omega}$ and $V_{xy}^{2\omega}$ at $D/\varepsilon_0 = 0.1$ Vnm$^{-1}$, at the high and low states, as a function of $D/\varepsilon_0|_{max}$, obtained using asymmetric sweep range. Here, $D/\varepsilon_0|_{min}$ was kept fixed at $-0.675$ Vnm$^{-1}$, and $D/\varepsilon_0|_{max}$ was varied. See SI section 17 for $D$ sweep range-dependent hysteresis from device s1.4. Clearly, the controllable hysteresis enables a second-order memory with multiple steady states, achievable by tuning the sweep range. This behaviour is analogous to electronic plasticity which stems from the flexoelectric mechanism in our tDBLG superlattice. Electronic plasticity has not been demonstrated in any single-element materials so far, as it requires polar elements or external charge trapping layers.

Now we demonstrate the synaptic response [51] of our second-order memory device, enabled by the $D$ sweep range-dependent plasticity discussed above (from sample s0.7, at $I_{ds}^{\omega} = 200$ nA, $n \approx 10^{12}$ cm$^{-2}$). In a biological synapse, the application of a pre-synaptic pulse, decreases or increases the synaptic conductivity/strength. In general, a particular state of synaptic conductance is achieved by using a sequence of pulses. An increase and decrease in synaptic strength are termed potentiation and depression, respectively. To achieve this functionality, we employ a sequence of fixed (– 0.675 Vnm$^{-1}$) negative triangular pulse of a $D/\varepsilon_0|_{min}$ followed by a tunable positive triangular pulse of $D/\varepsilon_0|_{max}$. For both types of pulses, the $|D/\varepsilon_0|$ was swept at a rate of 0.05 Vnm$^{-1}$s$^{-1}$. Figure 5a shows a typical depression cycle. The top and the middle panels show the time-series data for $V_{xy}^{2\omega}$ and $V_{xx}^{2\omega}$, while applying a sequence of pulses (shown in the bottom panel). The fixed negative reset pulse brings the memory to the low state. A tunable high state can be achieved simply by controlling the height of the positive set pulse. The data also demonstrate that the high states are stable over a long time. Figure 5b shows the potentiation cycle where the magnitudes of $V_{xy}^{2\omega}$(top panel) and $V_{xx}^{2\omega}$(middle panel) were decreased by gradually decreasing the height of the positive pulses. The bottom panel demonstrates the time-series data for the pulses. The height of the positive ($D/\varepsilon_0|_{max}$) pulses was varied from 0.675 Vnm$^{-1}$ to 0.2 Vnm$^{-1}$ for the potentiation cycle and from 0.2 Vnm$^{-1}$ to 0.675 Vnm$^{-1}$ for the depression cycle. In Figure 5c we demonstrate the long-term plasticity of the second-order response by showing several depression and potentiation cycles marked with horizontal black and red arrows, respectively. See SI section 16 for the data on the first-order response $V_{xx}^{\omega}$. Also see SI section 17 for the first and second-order synaptic response from device s1.4. The second-order voltage response ( $V_{xx}^{2\omega}$ and $V_{xy}^{2\omega}$ ) is pronounced near the integer superlattice filling and changes sign (Figure 3d) when $n$ is tuned across the integer fillings ($n/n_s = 0, \pm1, \pm2, \pm3$). This leads to a reconfigurable second-order synaptic memory functionality where the sign and magnitude of the states can be tuned by choosing a suitable $n$ for the device operation.

Finally, we evaluate the figures of merit of our synaptic memory in SI Section 18. The multi-level synaptic states (>16) exhibit high stability and endurance over repeated cycling (>>100), along with long, transient-free retention times, showing decay rates as low as ~ 0.2% per day. By inducing synaptic state changes using bipolar pulses of 2 ms duration, we calculate energy ($E$) consumption per synaptic event as $E \approx 0.5 - 0.8$ pJ (at $I_{ds}^{\omega} = 200$ nA), which is on per with state-of the-art first order synaptic transistors [13]. $E\ (\propto {I_{ds}^{\omega}}^2)$ can be further reduced drastically by operating at lower $I_{ds}^{\omega}$.

## 3. Conclusion

In conclusion, we demonstrate that strained moiré superlattices of twisted double bilayer graphene display robust hysteresis and electronic plasticity under the application of cyclic vertical displacement field. Such hysteresis does not require any polar elements or charge-trapping levels in the system. The hysteresis can be precisely controlled by adjusting the sweep range of vertical displacement field and persists up to 100 K. Combining the electrical hysteresis with the second-order nonlinear electrical response inherent to this inversion symmetry-broken system we develop a second-order synaptic memory device, demonstrating long-term synaptic plasticity. The multi-level synaptic states can be reconfigured in both sign and magnitudes using the naturally high electrostatic tunability of the second-order nonlinear electrical response in moiré superlattices. Such second-order synaptic memory could be crucial for low-power neuromorphic applications.

## 4. Experimental section

**Device Fabrication:** Hexagonal boron nitride (hBN) and bilayer graphene (BLG) crystals were mechanically exfoliated onto clean $SiO_2$ (300 nm ± 5%) / p++-Si substrates. hBN flakes (15–40 nm thick) were selected for use as encapsulation and dielectric layers. Heterostructures were assembled via a dry transfer method using a robotic manipulator-equipped, Ar-filled glovebox. The pickup-and-stack technique was employed to construct the device. To create a twisted double bilayer graphene (tDBLG) superlattice, the BLG flake was mechanically torn using a sharp tip. Detailed fabrication steps, including the tear-and-stack process, are provided in Supporting Information (SI) Section 1. **Electrical Measurements:** hBN served as the top-gate dielectric, while a 300 nm $SiO_2$ layer acted as the bottom-gate dielectric. Carrier density ($n$) was calculated as: $n = (\epsilon_0/e)[(\epsilon_{tg}(V_{tg} - V_{t0})/d_{tg}) + (\epsilon_{bg}(V_{bg} - V_{b0})/d_{bg})]$ , where $V_{tg}$ and $V_{bg}$ are top- and bottom-gate voltages, $\epsilon_{tg}$ ( $d_{tg}$ ) and $\epsilon_{bg}$ ( $d_{bg}$ ) are the dielectric constants (thicknesses) of the top and bottom dielectrics, and $V_{t0}$ and $V_{b0}$are gate voltage offsets at the charge neutrality point (CNP). The top-gate dielectric thickness ($d_{tg}$) was determined via progression of CNP in the $V_{tg}-V_{bg}$ phase space of sheet resistance ($R_S$) and validated by Hall measurements. The displacement field ($D$) is determined by: $2D/\epsilon_0 = [(\epsilon_{tg}(V_{tg} - V_{t0})/d_{tg}) - (\epsilon_{bg}(V_{bg} - V_{b0})/d_{bg})]$, where e is the electron charge and $\epsilon_0$ is vacuum permittivity. We simultaneously vary both $V_{tg}$ and $V_{bg}$ to keep one of $n$ and $D$ constant and vary the other. A high series-resistance ($\gg R_S$) is used to convert the lock-in reference out ( $V_{Out}^{\omega}$ ) into a current source ( $I_{ds}^{\omega}$ ). Synchronized lock-in amplifiers (17.77 Hz; input impedance $\gg R_S$ ) were used for simultaneously measuring both first ($\omega$) and second order ($2\omega$) voltage drops in longitudinal ($V_{xx}^{\omega}, V_{xx}^{2\omega}$) and transverse ($V_{xy}^{\omega}, V_{xy}^{2\omega}$) direction. Dual-channel DC source meters applied gate voltages. Measurements were conducted in a closed-cycle cryostat across 1.8–300 K under vertical magnetic fields of ±9 T.

**Supporting information**

The Supporting information (SI) is available. It contains the following sections. Section 1: Device details and fabrication. Section 2: Raman Characterization. Section 3: Evidence of twist angle anomaly from sample s1.3. Section 4: Landau Fan diagram. Section 5: $n - D$ phase space of doubly split $R_S$ maxima. Section 6: $D$ dependent hysteresis in the samples s1.3, s1.4, s1.4(2), s1.5, s1.9. Section 7: Stability of the hysteresis across several $D$ cycles. Section 8: $D$ sweep range dependence. Section 9: Hysteresis at different integer superlattice fillings. Section 10: $T -$ dependence of hysteresis. Section 11: Absence of hysteresis in encapsulated BLG and tDBLG sample (s10) with $\theta \approx 10°$. Section 12: Shift in CNP. Section 13: Temperature-dependent scaling behavior. Section 14: Hysteresis in $V_{xy}^{2\omega}$. Section 15: Sweep range dependence of first and second-order hysteresis for symmetric and asymmetric sweep range. Section 16: Synaptic functionality in the first-order. Section 17: First and second-order synaptic functionality from sample s1.4. Section 18: Evaluation of figures of merit.

**Acknowledgements**

The authors acknowledge financial support from MICIU/AEI/10.13039/ 501100011033 (Grant CEX2020-001038-M), from MICIU/AEI and ERDF/EU (Projects PID2021-122511OB-I00 and PID2021-128004NB-C21),

from MICIU/AEI and European Union NextGenerationEU /PRTR (Grant PCI2021-122038-2A) and from the European Union (Project 101046231-FantastiCOF). T.A. acknowledges funding from the European Union under Marie Sklodowska-Curie grant agreement number 101107842 (ACCESS). K.W. and T.T. acknowledge support from the JSPS KAKENHI (Grant Numbers 21H05233 and 23H02052) and World Premier International Research Center Initiative (WPI), MEXT, Japan. **Author contributions:** T.A. and L.E.H. conceived the project. T.A. fabricated the samples, performed the measurements, and wrote the manuscript. T.T. and K.W. provided the hBN crystals. All authors discussed and commented on the paper.

**Conflict of interest**

The authors declare that they have no conflict of interests.

**Data availability statement**

All data needed to evaluate to the conclusions in the paper are present in the paper and/or the supporting information. All source data related to the findings in this study can be obtained from the authors. All the experimental data will be uploaded to the Zenodo repository upon publication.